\documentstyle[12pt,epsfig]{article}

\begin{document}
\title{Analytical estimation of the maximal Lyapunov exponent 
         in oscillator chains}
\author{Thierry Dauxois$^{1,2}$\thanks{Permanent address:
Laboratoire de Physique, 
ENS Lyon, URA-CNRS 1325, 69364 Lyon C\'{e}dex 07, France}, 
Stefano Ruffo$^1$\thanks{INFN, Firenze (Italy)} 
and Alessandro Torcini$^1$\thanks{INFM, Firenze (Italy)} \\
$^1$ Dip. di Energetica "S. Stecco", via S. Marta, 3, I-50139 Firenze,
         Italy\\
$^2$ Scripps Institution of Oceanography, 
University of California at San Diego,\\ La Jolla CA 92093-0230, USA }
\date{\today}
\maketitle
\begin{abstract}
An analytical expression for the maximal Lyapunov exponent
$\lambda_1$ in generalized Fermi-Pasta-Ulam oscillator chains
is obtained. The
derivation is based on the calculation of modulational instability
growth rates for some unstable periodic orbits. The result is compared
with numerical simulations and the agreement is good over a wide
range of energy densities $\varepsilon$. At very high energy density the
power law scaling of $\lambda_1$ with $\varepsilon$ can be also obtained
by simple dimensional arguments, assuming that the system is ruled
by a single time scale. Finally, we argue that for repulsive and
hard core potentials in one dimension $\lambda_1 \sim \sqrt{\varepsilon}$ 
at large $\varepsilon$.
\end{abstract}
\section{INTRODUCTION}

Many theoretical and numerical studies have been devoted
to the characterization of chaos in high-dimensional systems.
Nevertheless, several fundamental issues are not understood.
In particular, the relation between Lyapunov instability analysis
and phase space properties like diffusion of orbits, relaxation 
to equilibrium states, spatial development of instabilities 
remains to be clarified~(see Ref.~\cite{lichtlieb} for a review).

In this paper we present an analytical estimate of the largest 
Lyapunov exponent $\lambda_1$ of the Fermi-Pasta-Ulam (FPU) model~\cite{ford}
based on the study of modulational instabilities of linear waves.
The FPU model has played for dynamical system theory a similar
role to the Ising model in statistical mechanics. Let's just
quote some major studies and discoveries motivated by the FPU
numerical experiment: the introduction of the concept of
soliton~\cite{zabuskykruskal}, the prediction of the transition
to large-scale chaos by the resonance overlap criterion~\cite{izrailev},
the use of KAM perturbation theory and Nekhoroshev stability
estimates~\cite{galgani}, the numerical detection of the strong
stochasticity threshold~\cite{bocchieri}.

Some attempts already exist~\cite{eckmann,casetti} of computing
analytically $\lambda_1$ in the FPU model. Here we present a
completely different approach, which emphasizes the relevant
role played by some unstable periodic orbits related to the 
representation in Fourier modes~\cite{poggi-ruffo} 
(preliminary results were already presented in Ref.~\cite{noi}).

In the present contribution, our analysis will be limited to the 
asymptotic state, where energy is equally shared among all normal
modes.
In Sect.~2, the modulational instability of a plane wave on the lattice 
is discussed, while in Sect.~3 an approximate analytical expression for
$\lambda_1$ is obtained from the growth rates of the modulational
instability, and  is compared with numerical results.  Finally, 
in Sect.~4 it is shown that, for sufficiently high energy density
$\varepsilon$, the dynamics of the system, in the phase as well
as in the tangent space, is ruled by
a single time scale $\tau(\varepsilon)$ and some extensions of the
results to repulsive and hard core systems are presented.

\section{MODULATIONAL INSTABILITY ANALYSIS}

\noindent
Denoting by $u_n(t)$ the position of the $n$-th
atom ($ n \in \{1, \dots, N \}$), the equations of motion for the 
generalized FPU chain read
\begin{eqnarray}
\ddot{u}_n = u_{n+1} + u_{n-1} - 2u_n +
\beta\left[(u_{n+1}-u_n)^{2p+1} - 
(u_n - u_{n-1})^{2p+1}\right]
\label{sub}
\end{eqnarray} 
where $p$ is an integer with $p \ge 1$, the parameter $\beta$ is
fixed to $0.1$ for comparison with previous results 
and periodic boundary conditions have been adopted.
For sake of simplicity, we first report the analysis for $p=1$ 
(i.e. for the usual $\beta$-FPU model) and then we generalize 
the results to any $p$-value.

\noindent
Due to the periodic boundary conditions, the normal modes
associated to the linear part of Eq.~(\ref{sub})
are plane waves of the form
\begin{equation}
u_n(t) =\phi_0\left(e^{i\theta_n (t)}+e^{-i\theta_n (t)}\right)
\label{plane}
\end{equation}
where $\theta_n(t) = qn-\omega t$ and $q=2\pi k/N$.
The dispersion relation is $\omega^2(q)=4(1+\alpha)\sin ^{2} (q/2)$,
where $\alpha=12\beta\phi_0^2 \sin^2(q/2)$ takes into account the 
nonlinearity~ \cite{pietro}.
The modulational instability of such a plane wave 
is investigated by studying the linearized equation associated to the
envelope of the carrier wave (\ref{plane}).
Therefore, one introduces infinitesimal
perturbations in the amplitude and phase and looks for solutions
\begin{eqnarray} 
u_n(t) &=& \phi_0[ 1 + b_n (t) ]\ e^{i\left[\theta_n (t)+\psi_n(t)\right]}+
\phi_0[ 1 + b_n (t) ]\ e^{-i\left[\theta_n (t)+\psi_n(t)\right]}\nonumber\\
&=&2\phi_0[1+b_n(t)]\ \cos[qn-\omega t+\psi_n(t)]
\label{az}
\end{eqnarray}
where  $b_n$ and $\psi_n$ are assumed to be small 
in comparison with the parameters of the carrier wave.
Substituting Eq.~(\ref{az}) into the equations of motion and
keeping the second derivative (at variance with what has been
done for Klein-Gordon type equation~\cite{KivsharPeyrard,daumont}),
we obtain for the real and imaginary part of the secular term 
$e^{i(qn-\omega t)}$
\begin{eqnarray}
-\omega^2b_n+2\omega\dot\psi_n+\ddot b_n&=&(1+2\alpha)
\left[\cos q\,(b_{n+1}+ b_{n-1})-2b_n \right]\nonumber \\
&-&\alpha\left(b_{n+1}+ b_{n-1}-2b_n\cos q \right)
-(1+2\alpha)\sin q\,(\psi_{n+1}-\psi_{n-1}) \\
-\omega^2\psi_n-2\omega\dot b_n+\ddot \psi_n&=&
(1+2\alpha)\left[\cos q\, (\psi_{n+1}+\psi_{n-1})-2\psi_n\right]
\nonumber \\
&+&(1+2\alpha)\sin q\,(b_{n+1}-b_{n-1})+\alpha
\left(\psi_{n+1}+ \psi_{n-1}-2\psi_n\cos q  \right)
\end{eqnarray}

\noindent
Assuming further $b_n=b_0 \ e^{i(Qn-\Omega t)}+{\rm c.c.}$
and $\psi_n=\psi_0 \ e^{i(Qn-\Omega t)}+{\rm c.c.}$ 
we obtain  the two following equations for the secular term
$e^{i(Qn-\Omega t)}$
\begin{eqnarray}
b_0\Bigl[\Omega^2+\omega^2+2(1+2\alpha)(\cos q \cos Q -1)
-2\alpha(\cos Q-\cos q) \Bigr]
-2i\psi_0\left[\omega\Omega+(1+2\alpha)\sin q\sin Q \right]=0\\
\psi_0\Bigl[ \Omega^2+\omega^2+2(1+2\alpha)(\cos q\cos Q-1)
+2\alpha(\cos Q-\cos q)\Bigr]
+2ib_0\left[ \omega\Omega +(1+2\alpha)\sin q\sin Q\right]=0
\end{eqnarray}
Non trivial solution for $(b_0, \psi_0)$ are found only if
the determinant vanishes, i.e. if the following equation is
fullfilled:
\begin{eqnarray}
\Biggl[(\Omega+\omega)^2-4(1+2\alpha)\sin^2\left({q+Q\over2}\right)\Biggr]&&
\Biggl[(\Omega-\omega)^2-
4(1+2\alpha)\sin^2\left({q-Q\over2}\right)\Biggr]  \nonumber \\
&=& 4\alpha^2\left(\cos{Q}-\cos q\right)^2
\label{relatdispercorr}
\end{eqnarray}

\noindent
This equation admits 4 different solutions once $q$
(wavevector of the unperturbed  wave) and $Q$ (wavevector of the perturbation)
are fixed.  If one of the solutions is complex we have an instability
of one of the modes $(q\pm Q)$ 
with a growth rate equal to the imaginary part of the solution. Therefore,
one can derive the instability threshold for any initial
linear wave, i.e. any wavevector and any amplitude.

\noindent
For example, for 
$q=0$, we obtain $\Omega=\pm \sin\left({Q/ 2}\right)$, which
proves that the solution is obviously stable
since the zero-mode, corresponding to translation invariance,
is completely decoupled from the others.

\noindent
For $q=\pi$, one can easily see that Eq. (\ref{relatdispercorr}) admits
two real and two imaginary solutions if and only if
\begin{equation}
\cos^2{Q\over 2}>{1+\alpha\over 1+3\alpha}
\label{acrit}
\end{equation}

\noindent
Being $\alpha=12\beta\phi_0^2=3\beta(2\phi_0)^2=3\beta a^2$
(where $a= 2 \phi_0$),
this formula is equivalent to the formula found by Sandusky and
Page~\cite{sandusky}. Moreover, it can be easily shown that 
the first unstable mode, associated to the perturbation,
corresponds to $Q={2\pi/ N}$ and
therefore the critical amplitude $a_c$ above which 
the $\pi$-mode looses stability is:
\begin{equation}
a_c= \left[ {\sin^2\left({\pi\over N}\right)\over 3\beta
\left[3\cos^2\left({\pi\over N}\right)-1\right]} \right]^{1/2}
\end{equation}

\noindent
Since for the $\pi$-mode the energy is  simply given by 
$E=N(2a^2+4\beta a^4)$, we finally get the critical energy
\begin{equation}
E_c={2N\over 9\beta}\sin^2\left({\pi\over
N}\right){7\cos^2\left({\pi\over N}\right)-1
\over \left[3\cos^2\left({\pi\over N}\right)-1\right]^2}
\label{eqnc}
\end{equation}
For large values of $N$, we obtain,
in agreement with previous approximate estimations
\cite{poggi-ruffo,berman,flach}, the expression 
\begin{equation}
E_c={\pi^2\over 3\beta N}+O\left({1\over {N^3}}\right).
\end{equation}
that turns out to be quite accurate for $ N > 50 - 60$.
Above this energy threshold, the $\pi$-mode is therefore unstable
and gives rise to a ``chaotic breather'', that has a
finite life-time and finally disappears leading to energy
equipartition (this complex relaxation process to equipartition
is described in detail in Ref.~\cite{cretegny}).

\section{GROWTH RATES AND LYAPUNOV EXPONENTS}

\noindent
Again for the $\pi$-mode, from Eq.~(\ref{relatdispercorr}),
one can compute the growth rate
$\sigma(\pi,Q)={\rm Im}(\Omega(Q))$ from Eq.~(\ref{relatdispercorr})
\begin{equation}
\sigma(\pi,Q)=  2\sqrt{\sqrt{4(1+\alpha)(1+2\alpha)\cos^2{Q\over 2}
+\alpha^2\cos^4{Q\over 2}}-1-\alpha-(1+2\alpha)\cos^2{Q\over 2} }
\label{growthrate}
\end{equation}

\begin{figure}
\centering\epsfig{file=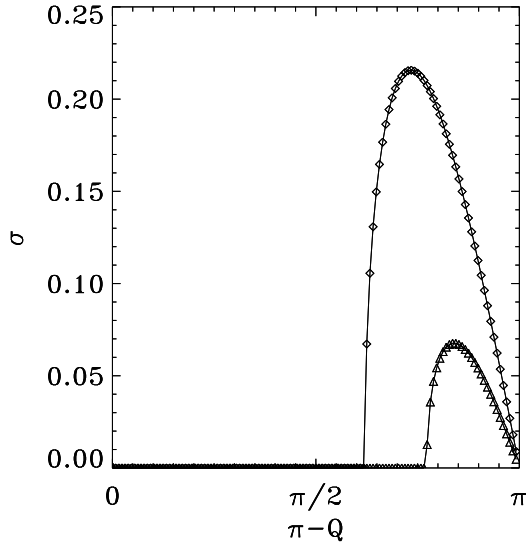,width=0.45\linewidth}
\caption{Shape of the growth rate $\sigma(\pi,Q)$. The diamonds
correspond to an amplitude $a=1$ whereas the triangles
to $a=0.5$.}
\protect\label{growth}
\end{figure}

\noindent
It is plotted in Fig.~\ref{growth} for two different amplitudes.
When the amplitude (or energy density) increases, the region of
instability extends to a larger region of wavevectors, and
in particular the most unstable mode $Q_{\hbox{max}}$
increases reaching an asymptotic value 
$
\tilde Q_{\max}=2\arccos\left(\sqrt{{8\over\sqrt3}-4}\right)\simeq 0.42 \pi
$
It is important to notice that,
for sufficiently high energy, the rescaled growth rate
$\sigma/\sigma(\pi,Q{\hbox{max}})$ does not depend on the energy density. 
This suggests that in the high energy limit a unique time
scale is present in the system; we will discuss in more detail this
point in the following.

\begin{figure}
\centering\epsfig{file=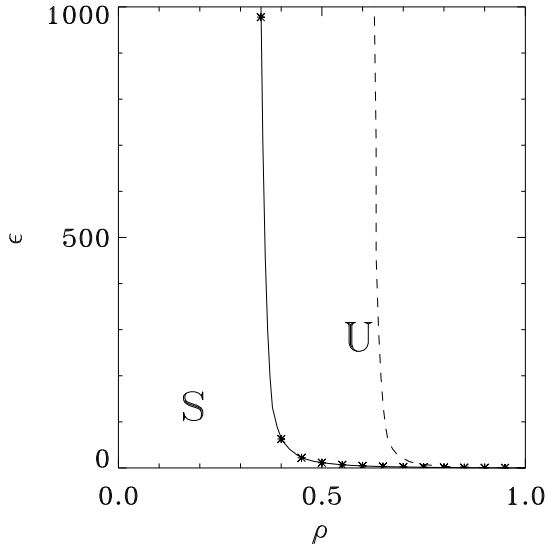,width=0.45\linewidth}
\caption{The solid line corresponds
to the  marginal stability curve $\varepsilon_0$,
the dashed one to $\varepsilon_M = \varepsilon(\rho_{\max})$
and the symbols to the points analytically found by Poggi and Ruffo 
\protect\cite{poggi-ruffo}.
The symbols $S$ and $U$ denote stable and unstable region, respectively,
separated by the solid line.}
\protect\label{pietro}
\end{figure}

\noindent
One can also derive the lowest unstable mode, whose asymptotic limit 
at high energy is
$\tilde Q_{min}=2\arccos{(1/ \sqrt 3)} \simeq0.6\pi\simeq 1.91 $
(as previously derived in \cite{poggi-ruffo}).
Fig.~\ref{pietro} reports the stability and instability regions
in the plane of the parameters $\varepsilon = E/N$
and $\rho(Q)=\cos^2(Q/2)$. In particular,
the marginal stability line $\varepsilon_0$, associated to 
the vanishing of $\sigma(\pi,Q)$, and the line $\varepsilon_M
= \rho(Q_{max})$ are reported.
For the marginal stability line $\varepsilon_0$, our results are
fully consistent with a previous theoretical estimation reported 
in~\cite{poggi-ruffo}.

\noindent
The above results can be generalized to any power $p$.
In particular, limiting the analysis to the $\pi$-mode, 
the instability condition (\ref{acrit}) takes the form
\begin{equation}
\cos^2{Q\over 2}>{1+\alpha\over 1+(2p+1)\alpha}~,
\label{acritnew}
\end{equation}
where $\alpha$ is now given by
$
\alpha=\beta \frac{(2p+1)!}{p ! (p+1) !}
\left(2\phi_0\sin{q\over 2}\right)^{2p}.
$
One can therefore derive the critical amplitude above which the
$\pi$-mode is unstable and for large $N$ we have
\begin{equation}
a_c\simeq \left[{\pi^2 p ! (p+1) ! \over 2p\beta N^2 (2p +1) !}
\right]^{1\over
2p}\propto N^{-{1\over p}}
\end{equation}
It means that (for fixed $N$) $a_c$ is an increasing function 
of the power of the coupling potential with asymptotic limit 
$\lim_{p\rightarrow\infty} a_c=0.5$. Therefore, in the hard potential
limit the critical energy density for the $\pi$-mode is
finite ($\varepsilon_c = 0.5$) for any finite $N$-value.
It should be noticed that this is not in contradiction with
the integrability of a one-dimensional system of hard rods,
because in the present case we have also a harmonic contribution 
at small distances.

\noindent
Finally, we observe that, for the generalized FPU-model, 
in the high energy limit the growth rate 
$\sigma_p(\pi,Q)\approx\sqrt\alpha\ \overline\sigma_p(\pi,Q)$,
where $\overline\sigma_p(\pi,Q)$ is independent of the energy.
In summary, for high enough values of the energy,
\begin{equation}
\sigma_p(\pi,Q)\propto \varepsilon^{{1\over 2}-{1\over 2(p+1)}}.
\label{scal}
\end{equation}
We will see in the last section that this scaling law can
be also derived from simple dimensional arguments. 

\begin{figure}
\centering\epsfig{file=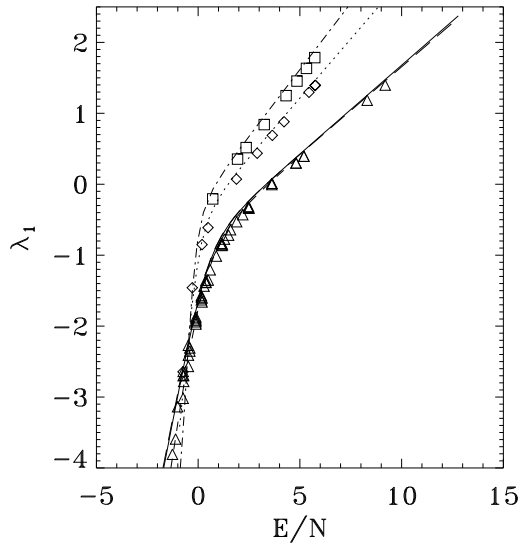,width=0.45\linewidth}
\caption{Comparison of the analytical estimate  with numerical results
for the maximal Lyapunov exponent. The solid curve corresponds to our 
estimation (Eq.~{\protect \ref{analexpr}}), the dashed
curve to the CLP-estimate using Riemannian differential geometry
(see Ref.~{\protect \cite{casetti}}) and the triangles to our 
numerical results (for the $\beta$-FPU, i.e. $p=1$).
The dotted curve (resp. diamonds) corresponds to the analytical 
estimate (resp. numerical results) for $p=2$ and the dash-dotted (resp.
squares)
 to the analytical estimate (resp. numerical results) for
$p=3$.
}
\protect\label{numtheo}
\end{figure}

\noindent
Let us now report an analytical estimation of the maximal Lyapunov.
As the system is Hamiltonian, Pesin's theorem allows us to identify the
Kolmogorov-Sinai entropy $S_{KS}$ with the sum
of all positive Lyapunov exponents. 
As the Lyapunov spectrum was shown~\cite{LPR} to be 
approximately linear at high energy,
one can relate $S_{KS}$ to the maximal Lyapunov exponent $\lambda_1$, namely
\begin{equation}
S_{KS}=\sum_{i=1}^N\lambda_i\cong{\lambda_1 {N\over 2}} \quad .
\end{equation}
Let us define a new quantity : the instability entropy
\begin{equation}
S_{IE}(q)=\sum_{i=1}^{N/2}\sigma(q,2\pi i/N)\quad,
\end{equation}
where the sum is over all positive growth rates~\cite{posgrowth}.
Our crucial physical {\it ansatz} is that $S_{KS} \simeq S_{IE}(\pi)$.
From this assumption the following expression for the
maximal Lyapunov exponent is readily derived:
\begin{equation}
\lambda_1={2\over N} S_{IE}(\pi)  \label{analexpr}\quad .
\end{equation}
Employing the analytical values for $\sigma(\pi,Q)$ reported in Eq.~(\ref{growthrate}), 
$\lambda_1$ can be finally computed.
Fig.~\ref{numtheo} attests that the analytical expression~(\ref{analexpr})
is very accurate. In the same figure the data obtained with a 
completely different approach, developed by Casetti, Livi and Pettini
(CLP)~\cite{casetti}, are also shown. 
The two methods give almost identical results, apart at
very low energy, where the CLP findings~\cite{casetti} is 
in better agreement with our numerical data \cite{noi}. In particular,
the low energy scaling turns out to be $\lambda_1 \sim \varepsilon^{2}$, 
instead our formula gives $\lambda_1 \sim \varepsilon^{1.5}$.

\noindent
As a matter of fact, we obtain a very good agreement 
between our theoretical
estimation and the numerical results also for $p=2$ and 3 
(see Fig.~\ref{numtheo}).
Plotting  $\lambda_1 (E/N)$ in a log-log scale we observe
(see Fig.~\ref{numtheo}) that at high energy
an asymptotic linear behaviour is found for $p=1,2$
and $3$. In the low energy limit, however, 
for any $p$ the same scaling behaviour should be found,
as confirmed by the data reported in Fig.~\ref{numtheo}.
The two asymptotic linear behaviors (at high and low
$\varepsilon$) are separated by a knee at intermediate $\varepsilon$. 
An estimation of this transition value can be obtained assuming that the
linear and nonlinear contributions to $\omega(q)$ should
be of the same order. We obtain $\alpha \sim 1$ i.e.,
an energy density of the order of $1/\beta$
(this is equivalent to the estimation given 
by mode overlap criterion\cite{izrailev}).
This knee corresponds to a stochasticity
threshold~\cite{bocchieri} which defines the crossing from weak to
strong chaos and has been called strong stochasticity threshold (SST) 
(similar SST-transitions have been recently
identified also for other models of nonlinear chains \cite{yoshi}).
However, we have also found that it corresponds
to a crossover from extended to a more localized state 
in tangent space \cite{noi}.

\noindent
It is remarkable to note that  Chirikov~\cite{chirikov}
found similarly the maximal Lyapunov exponent
of the standard map at high energy
by averaging over the phase space the maximal eigenvalue
associated to the main hyperbolic point. It corresponds
in our case to averaging the growth rate~(\ref{growthrate})
for the unstable periodic orbit $q=\pi$ over the equilibrium 
equipartition state (where all modes have the same weight).
A similar approach is known as Toda criterion~\cite{benettingal},
and although it cannot be used as a signature of chaos, it can 
give an approximate estimation of $\lambda_1$.

\noindent
In fact, one can understand this average in a better
way by recalling that the modes $\{\pi/2\}$,$\{2\pi/3\}$,$\{\pi\}$
correspond to the simplest unstable periodic orbits and 
are also the only three one-mode solutions of the $\beta$-FPU
problem \cite{poggi-ruffo}. The calculation of the 
instability entropies of this three modes shows that they are
extremely close one to another, contrary to 
the value for other modes. Recently, it has been found that
very good estimates for $\lambda_1$ can be obtained by
applying the CLP method directly to any of the above three
one-modes solutions \cite{franz-poggi}. 
This confirms that the dynamics of these one-mode solutions is 
quite peculiar, because they seem to be "optimal" trajectories 
in the phase-space to estimate the maximal Lyapunov exponent.

\begin{figure}
\centering\epsfig{file=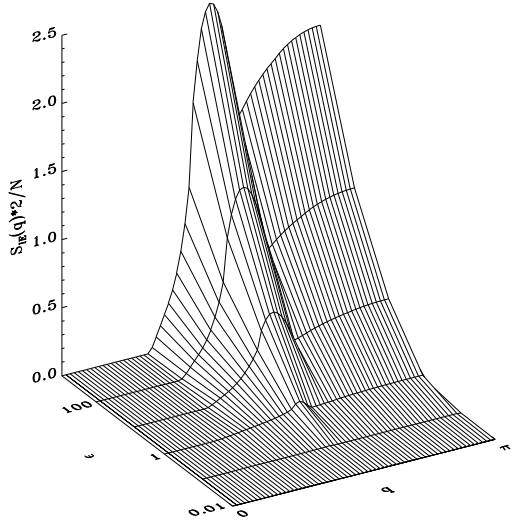,width=0.45\linewidth}
\caption{Normalized instability entropy $2S_{IE}/N$ as a function
of $q$ and $\varepsilon$.
}
\protect\label{yoshi}
\end{figure}

\noindent
The normalized instability entropy $2S_{IE}(q)/N$ is plotted against $q$
and $\varepsilon$ in Fig.~\ref{yoshi}. This should be compared with
Fig.~2a in Ref.~\cite{yoshi0}, where a similar plot is
reported for the Lyapunov exponent computed from the true
Jacobian matrix, but taking the linear Fourier modes as
approximate orbits (moreover the case of fixed boundary
conditions is considered). At fixed $\varepsilon$ both these
quantities initially grow with $q$ (low $q$'s are more stable 
than high $q$'s), but then a ``stability'' region (meaning that
the growth rate of the instability is significantly smaller)
is present around $q=2\pi/3$ (numerically estimated as 0.7 $\pi$
by Yoshimura in Ref.~\cite{yoshi0}). The ``ridge'' structure 
observed in~\cite{yoshi0} reduces in our case to two peaks. The
location of the first one depends on $\varepsilon$ and reaches
$q\approx 0.562\, \pi$ for large energy density (it is around
$0.5 \pi$ in the region explored by Yoshimura). The second
peak is at $(N-2)\pi / N $, and thus goes to $\pi$ in the 
$N \to \infty$ limit.
We remark that our evaluation of the instability takes fully
into account nonlinear effects, contrary to Yoshimura's, and that
it is completely analytical (apart from the necessary numerical 
evaluation of the roots of Eq.~(\ref{relatdispercorr})).
Mainly, we can confirm that a ``stability'' region 
(in the previous sense) is present around $q=2\pi/3$, 
which corresponds to one
of the exact solutions found in Ref.~\cite{poggi-ruffo}. 

\section{HIGH ENERGY LIMIT}

\noindent
Here, our claim is that in the high energy limit the temporal evolution
of system (\ref{sub}) is essentially ruled by a single time-scale.
This time-scale can be easily derived by a simple
dimensional argument (see also~\cite{eckmann}).
At high energy density the potential energy per particle scales as
$
\frac{V}{N} \sim u^{2p+2}~,
$
where $u$ is the typical size of $u_n$, and similarly for the kinetic 
energy per particle
$
\frac{K}{N} \sim \dot{u}^2~.
$
Assuming a virial of kinetic and potential energies (this is shown to hold
for the FPU model in Ref.~\cite{casartelli})
$
\frac{V}{N} \sim \frac{K}{N} \sim \frac{\varepsilon}{2}~,
$
we can introduce a typical time scale
\begin{equation}
\tau \sim \frac{u}{\dot{u}} \sim \varepsilon^{-\frac{1}{2}+\frac{1}{2(p+1)}}
\label{scaling}
\end{equation}
Therefore, the scaling reported in Eq.~(\ref{scal}) is easily recovered and
it is natural to expect that the same scaling should hold also for $\lambda_1$.
In particular, for $p=1$, $\lambda_1$ scales as $\varepsilon^{1/4}$
(as also found in Ref.~\cite{casetti}).

\noindent
A different scaling, with power 2/3, was found in Ref.~\cite{pettini}
at smaller energy densities (closer to the strong stochasticity
threshold (SST), the knee of Fig.~3). A random matrix 
approximation~\cite{cripalvul} was invoked to explain these results, 
which is legitimate only if the process is uncorrelated and one 
is approaching an integrable limit, where the Lyapunov exponent 
vanishes. Both these approximations cannot hold exactly in the region where 
the 2/3 exponent was found, although a fast decrease of the Lyapunov
exponent at the knee could mimick the vanishing exponent situation.

\noindent
For the $\beta$-FPU model ($p=1$) in the high energy limit, an identical 
scaling has been recently found by Lepri~\cite{lepri} for the
decay-rate of the time autocorrelation functions of the Fourier modes and
in Ref.~\cite{cretegny} for the inverse of the equipartition time.
We have also verified that the normalized Lyapunov spectrum 
$\tilde \lambda_i = \lambda_i/\lambda_1$ converges to an 
asymptotic shape (independent of $\varepsilon$) for $\varepsilon \to \infty$.
These results confirm our assumption that the dynamics of the 
system is essentially ruled by an unique time-scale 
for sufficiently high energy density.

\begin{figure}
\centering\epsfig{file=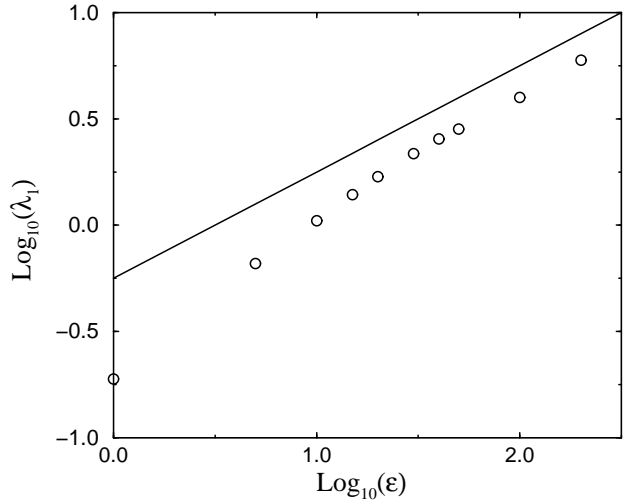,angle=-90,width=0.45\linewidth}
\caption{Logarithm of the maximal Lyapunov exponents for the TCL lattice model
as a function of the logarithm of the energy density $\varepsilon$.
Circles represent our numerical data obtained with $N=64$
and adopting a 6-th order symplectic integrator {\protect{\cite{yoshida}}},
with a relative energy conservation ranging from $10^{-12}$ to
$10^{-6}$. The solid line refers to a slope 0.5.}
\label{tcl}
\end{figure}

\noindent
It is important to stress that in the limit of hard core potentials
($p\rightarrow\infty$) we find $\lambda_1 \sim \varepsilon^{1/2}$.
Moreover, also for potentials that are repulsive (i.e. diverge 
as $(x-c)^{-q}$ for $x \to c$ with $q \ge 1$, or even steeper),
we argue that the maximal Lyapunov exponent scales as 
$\varepsilon^{1/2}$ in the high energy density limit.
This claim is also supported by the numerical results in Ref~\cite{yoshi},
where the author studies three different types of potentials,
with a repulsive part at short distances: a diatomic Toda lattice,
a truncated Coulomb potential (TCL) $V(x) = \frac{1}{2}[\frac{1}{x+1} +x-1]$
and a Lennard-Jones (LJ) potential.
For the TCL potential and the LJ potential, small deviations from
the 1/2 power are found by Yoshimura, but these tend to reduce as
the energy density increases, as shown by our simulations of the
TCL model reported in Fig.~\ref{tcl}.  
In Ref.~\cite{yoshi} also the equipartition times are reported, whose
scaling with $\varepsilon$ at high energy density is always consistent 
with an exponent $-1/2$. 

\noindent
Finally, for a gas of hard spheres in a three dimensional box 
the analogue of the energy
density is the temperature $T$. It is well known that for
this model the characteristic time is represented by the
Enskog collision time $\tau_E$, which for constant density 
is proportional to $T^{-1/2}$ \cite{baluc}. 
This result suggests that the scaling law $\tau \sim \varepsilon^{-1/2}$ 
should be valid also for a gas of hard spheres at constant density
even in three dimensions, as confirmed by recent data
for $\lambda_1$ \cite{posch}.

\section{CONCLUSIONS}
We have shown that the mechanism of modulational instability of a
plane wave (for the three admissible exact one-mode solutions) 
is strictly connected to Lyapunov instability and
its study gives quantitative formula for the maximal Lyapunov
exponent. This is still somewhat mysterious, because this 
mechanism deals with the short-time behavior, while the 
Lyapunov instability arises from an infinite-time average. One possible
explanation of the mystery is that the considered unstable
orbits perform a good enough sampling of the phase space, which
is uniformly visited by a generic orbit in the equipartition state.

Scaling laws for the Lyapunov exponent at high energy density for
generalized FPU chains have been obtained using the modulational
instability approach, but it is also observed that they can be
easily guessed by simple dimensional arguments. This points to
the existence of a universal time scale at high energy, independent
of the considered quantity (whether it is the Lyapunov exponent
or a correlation function). The same universality holds for
hard core potentials.

\section*{Acknowledgements} 

We would like to thank U. Balucani, R. Franzosi, S. Lepri, 
P. Poggi, A. Politi and A. Rapisarda for 
enlightening discussions as well as L. Casetti, R. Livi and M. Pettini, 
for providing us some of their data.  T.D. gratefully acknowledges EC
for financial support with grant ERBFMBICT961063.
This work is also part of the European contract
ERBCHRXCT940460 on ``Stability and universality in classical
mechanics''.  S.R. acknowledges the hospitality of the {\it Centro
Internacional de Ciencias} in Cuernavaca, M\'exico and the financial
support of the same Center and of CONACYT. 
Part of CPU time has been nicely supplied by the 
Institute of Scientific Interchanges (ISI) of Torino.

\noindent
Giovanni Paladin gave important contributions to the theory
of Lyapunov exponents, we think he would have been interested
in these results.


\begin{thebibliography}{99}
\bibitem{pietro} For $q = \pi/2$, $2 \pi /3$ and $\pi$, solution
(\protect{\ref{plane}}) is the first term of the
harmonic expansion of an exact solution of the $\beta$-FPU
model \cite{poggi-ruffo}.
\bibitem{posgrowth} For a given wavevector, there are 4 solutions of 
Eq.~(\ref{relatdispercorr}). Two are real and two complex conjugates:
i.e. 2 solutions with positive (or null) growth rates and two 
with negative (or null) growth rates.
\end{thebibliography}
\end{document}